# Superconducting Quantum Interference Devices based on InSb nanoflag Josephson junctions

Andrea Chieppa,[†] Gaurav Shukla,[†] Simone Traverso,[‡,¶] Giada Bucci,[†] Valentina Zannier,[†] Samuele Fracassi,[‡,¶] Niccolo Traverso Ziani,[‡,¶] Maura Sassetti,[‡,¶] Matteo Carrega,[*,¶] Fabio Beltram,[†] Francesco Giazotto,[†] Lucia Sorba,[†] and Stefan Heun[*,†]

[†]*NEST, Istituto Nanoscienze-CNR and Scuola Normale Superiore, Piazza San Silvestro 12, 56127 Pisa, Italy*

[‡]*Dipartimento di Fisica, Università di Genova, Via Dodecaneso 33, 16146 Genova, Italy*

[¶]*CNR-SPIN, Via Dodecaneso 33, 16146 Genova, Italy*

E-mail: matteo.carrega@spin.cnr.it; stefan.heun@nano.cnr.it

## Abstract

Planar Josephson junctions (JJs) based on InSb nanoflags have recently emerged as an intriguing platform in superconducting electronics. This letter presents the fabrication and investigation of superconducting quantum interference devices (SQUIDs) employing InSb nanoflag JJs. We provide measurements of interference patterns in both symmetric and asymmetric geometries. The interference patterns in both configurations can be modulated by a back-gate voltage, a feature well reproduced through numerical simulations. The observed behavior aligns with the skewed current-phase relations of the JJs, demonstrating significant contributions from higher harmonics. We explore the magnetic field response of the devices across a wide range of fields (±30 mT), up to the single-junction interference regime, where a Fraunhofer-like pattern is detected. Finally, we assess the flux-to-voltage sensitivity of the SQUIDs to evaluate their performance as magnetometers. A magnetic flux noise of $S_\Phi^{1/2} = 4.4 \times 10^{-6}$ $\Phi_0/\sqrt{\text{Hz}}$ is identified, indicating potential applications in nanoscale magnetometry.

Superconducting quantum interference devices (SQUIDs) are one of the most important classes of devices in quantum technologies. The fundamental component of a SQUID is the Josephson junction, which is created by sandwiching a normal (non-superconducting) material between two superconducting electrodes. When two Josephson junctions are connected in parallel to a loop, a dc-SQUID is formed.[1] This type of device is the most sensitive magnetometer,[2,3] with crucial applications in scanning probe microscopies[4–6] and superconducting electronics.[7,8] Furthermore, SQUIDs in highly asymmetric configurations, where one arm of the SQUID carries a supercurrent significantly larger than the other, have been explored as a way to investigate a fundamental property of Josephson junctions: the current-phase relation (CPR).[9–16] This key quantity is challenging to test directly in experiments.

With its small effective mass, narrow bandgap, and significant spin-orbit interaction, InSb has emerged as a highly sought-after material for applications in high-frequency electronics and spintronics.[17] Additionally, with a large Landè g factor, superconducting-semiconducting hybrids incorporating InSb have been proposed as a platform to host topological superconductivity and search for Majorana fermions,[18–21] motivating the investigation of this material in conjunction with superconductors. Recently, InSb nanoflags have surfaced as a promising platform featuring quasi-2D electronic transport. These InSb nanoflags, epitaxially grown by chemical beam epitaxy, are free-standing semiconducting nanostructures.[22–26] There, ballistic Josephson junctions with InSb nanoflags have been fabricated,[27] and non-reciprocal transport along with half-integer Shapiro steps has been reported.[28,29] A non-sinusoidal CPR was proposed to elucidate these observations. However, no direct measurement of this CPR has been reported, indicating that experimental studies remain extremely important. Among the consequences of a non-sinusoidal CPR, non-reciprocal transport in supercurrent interferometers, such as SQUIDs, has recently been at the center of several works.[30–39] Substantial higher-harmonic content and asymmetry between the two junctions forming the SQUID are necessary to observe this effect. An asymmetry between the two arms can be introduced by considering different geometrical aspect ratios (length/width) of the two Josephson junctions forming the SQUID. Furthermore, it can be induced by different interface transparencies, which would directly affect the skewness of the respective CPRs.[40]

In this letter, we have fabricated and investigated SQUIDs that incorporate Josephson junctions based on InSb nanoflags, aiming to examine the skewness and harmonic content inherent in these proximitized Josephson junctions. To achieve this, we developed SQUIDs in various geometrical configurations, scrutinizing their behavior under the influence of back-gate voltage variations and external magnetic fields. Additionally, we explore the potential applications of these hybrid structures in detecting minute magnetic signals, thereby evaluating their performance as magnetometers in nanoscale devices.

The SQUIDs investigated are based on Josephson junctions with InSb nanoflags as normal material and niobium (Nb) as superconducting material (measured $T_c = 8.1\,\text{K}$). The InSb semiconducting nanostructures possess a zinc blende crystal structure. They are typically $2.8 \pm 0.2\,\mu\text{m}$ long, $470 \pm 80\,\text{nm}$ wide, $105 \pm 20\,\text{nm}$ thick, and characterized by quasi-2D transport.[22] Control over the carrier density in InSb nanoflag-based Josephson junctions is obtained by capacitively coupling a p-type doped silicon back-gate through $285\,\text{nm}$ of $SiO_2$. In the depletion region, when the back-gate voltage $V_{\text{bg}}$ is below a threshold voltage $V_{\text{th}}$ ($V_{\text{bg}} < V_{\text{th}}$), the conductance of the junction is zero. Increasing the back-gate voltage opens the semiconducting channel, and a steep increase in the conductance is observed.

Because of their elongated shape, nanoflag-based Josephson junctions can be fabricated in different configurations. The "narrow" configuration (JJ1 in Fig. 1(b1)), extensively investigated in previous works,[27–29,41] is obtained by depositing the superconductor along the shorter direction of the nanoflags. This limits the contact width to $\sim 600\,\text{nm}$, but grants the possibility of tuning the spacing of the Nb-electrodes from micrometers ("Hall bar" or long junction regime) to tens of nanometers (short junction regime). A "wide" configuration (JJ2 in Fig. 1(b2)) is obtained by rotating the nanoflags 90 degrees, allowing for larger contact widths, up to 3 µm, with an electrode spacing of submicrometer size.

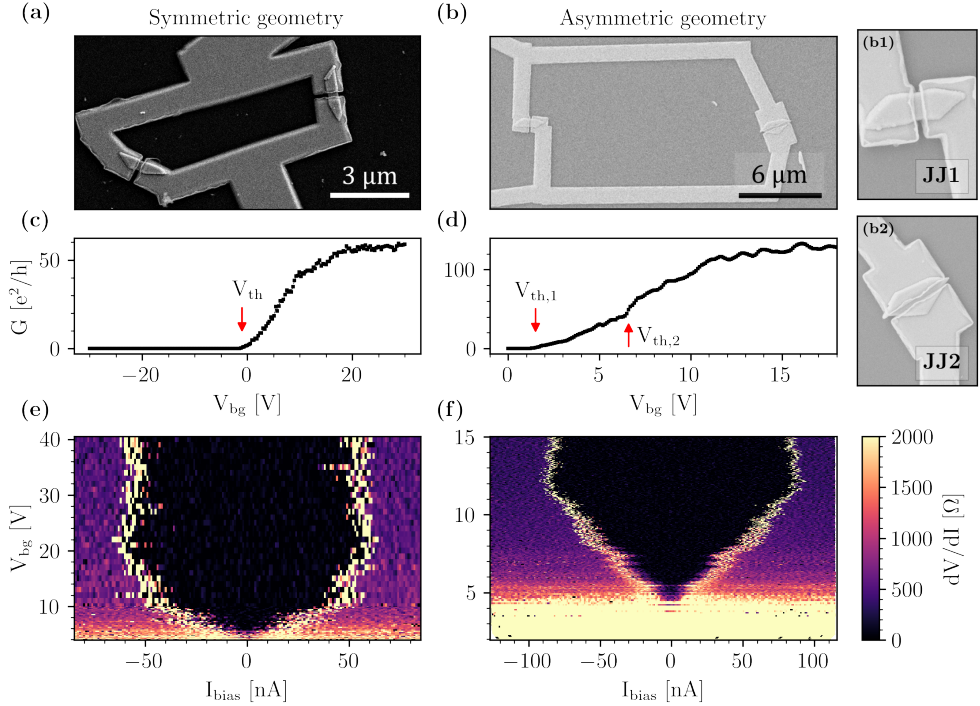

Figure 1: (a), (b) Top-view scanning electron microscopy images of the devices characterized in this work. (a) SQUID in a symmetric geometry. (b) SQUID in an asymmetric geometry. (b1), (b2) Enlarged views of the two Josephson junctions forming the asymmetric SQUID. (b1) Narrow configuration. (b2) Wide configuration. (c), (d) Conductance vs. back-gate voltage for (c) the symmetric and (d) the asymmetric geometry, measured at $T = 2\,\text{K}$. (e), (f) Color maps of the differential resistance ($dV/dI$), for the device in (a) and (b), respectively, as a function of back-gate voltage, obtained as the numerical derivative of the measured $V - I$ curves. $T = 350\,\text{mK}$.

This degree of freedom in the configuration is used to realize two SQUID geometries: symmetric and asymmetric. In symmetric geometry, both Josephson junctions have a narrow configuration, as shown in Fig. 1(a), with width $W_1 = W_2 = 380\,\text{nm}$. The total loop area is $A_\text{geo} = 13.6\,\text{µm}^2$. In the asymmetric geometry, shown in Fig. 1(b), instead, one junction has the wide configuration ($W_1 = 1.7\,\text{µm}$), while the other the narrow one ($W_2 = 530\,\text{nm}$), with area $A_\text{geo} = 118\,\text{µm}^2$. Additional information on fabrication methods, with a table of full geometrical parameters, is available in Section S1 of the Supporting Information (SI). Section S2 of the SI provides details on the measurement methods.

We first characterize the SQUIDs by investigating their transport properties at zero magnetic field. As SQUIDs consist of the electrical parallel of two Josephson junctions, the measured conductance can be expressed as the sum of each junction's conductance:

$$G_{\text{S}}(V_{\text{th},1}, V_{\text{th},2}) = G_{\text{JJ},1}(V_{\text{th},1}) + G_{\text{JJ},2}(V_{\text{th},2}), \tag{1}$$

where the voltage threshold $V_{\text{th,i}}$ of each junction has been introduced. In measuring the conductance vs. back-gate curves in the normal state for the symmetric geometry, a single threshold is found (red arrow, Fig. 1(c)). In contrast, in the asymmetric geometry, two different threshold values are extracted (red arrows, Fig. 1(d)). Hence, in the asymmetric SQUID, there is a range of back-gate voltages for which only one arm of the interferometer is conductive. Two distinct voltage thresholds can be attributed to fabrication inhomogeneities that would result in the conduction band edges of the two nanoflags not being perfectly aligned. This, in turn, would cause the chemical potential to be brought into the band gap at two different back-gate voltages for the two junctions.

Looking at superconductive properties, the Josephson effect at zero magnetic field is investigated by measuring voltage-current $(V - I)$ characteristics as a function of the back-gate voltage in a DC setup. At the temperatures explored here ($T = 350\,\text{mK}$ and above), no significant differences are observed between the switching and retrapping currents, which places these SQUIDs in the non-hysteretic regime. The differential resistance, obtained by numerical differentiation of the measured $V - I$ curves, is plotted against the voltage of the back-gate and the current bias in Fig. 1(e) and Fig. 1(f) for the symmetric and asymmetric device, respectively. The back-gate modulates the critical current in both configurations to pinch off, as expected.[27] Increasing $V_{\text{bg}}$, the critical current presents a non-monotonous behavior, reaching a maximum value of 20 V and 12 V for the symmetric and asymmetric case, respectively, and then slightly decreases. This modulation allows to exclude Nb accidental shorts or other transport channels different from the InSb nanoflags.

Before discussing the experimental results with an applied perpendicular magnetic field, it is convenient to introduce the key points of the SQUID interference. The total supercurrent flowing through a SQUID, $I_S$, is the sum of the supercurrents carried by each arm, and is a function of the superconducting phase drop across both junctions, $\varphi_i$,

$$I_\text{S}(\varphi_1, \varphi_2) = I_1(\varphi_1) + I_2(\varphi_2). \tag{2}$$

The superconducting phase drops $\varphi_1$ and $\varphi_2$ are linked via the flux quantization condition

$$\varphi_1 - \varphi_2 = 2\pi \frac{\Phi}{\Phi_0}, \tag{3}$$

where $\Phi_0 = h/2e \simeq 2.068\,\text{mT}\,\text{µm}^2$ is the superconducting flux quantum and $\Phi = \Phi_\text{ext} + LI_\text{circ}$ the total flux enclosed in the SQUID loop, which is the sum of the externally applied flux $\Phi_\text{ext}$ and an induced flux $LI_\text{circ}$. The induced flux is due to a supercurrent circulating in the loop, $I_\text{circ} = (I_1 - I_2)/2$, and the total inductance $L$, expressed in terms of a geometrical and a kinetic contribution. The impact of a finite inductance on the phase difference reported by Eq. 3 is given by the inductance parameter $\beta_\text{L} = 2\pi L I_\text{c}/\Phi_0$. In our case, considering that $L$ is of the order of $10\,\text{pH}$ (further details can be found in Section S3 of the SI) and $I_\text{c}$ of the order of $100\,\text{nA}$, we get $\beta_\text{L} \approx 10^{-3}$. Thus, corrections to the enclosed magnetic flux due to the self-inductance can be neglected. In the following we will consider $\Phi \approx \Phi_\text{ext}$. The critical current of the SQUID is obtained by maximizing the total supercurrent over the phase difference $\varphi_1$,

$$I_\text{c}(\Phi) = \max_{\varphi_1} I_\text{S}(\varphi_1, \Phi), \tag{4}$$

and is a function of the enclosed flux only. Thus, the shape and the features of the interference pattern $I_\text{c}(\Phi)$ depend directly on the CPRs $I_\text{i}(\varphi)$ of the two Josephson junctions forming

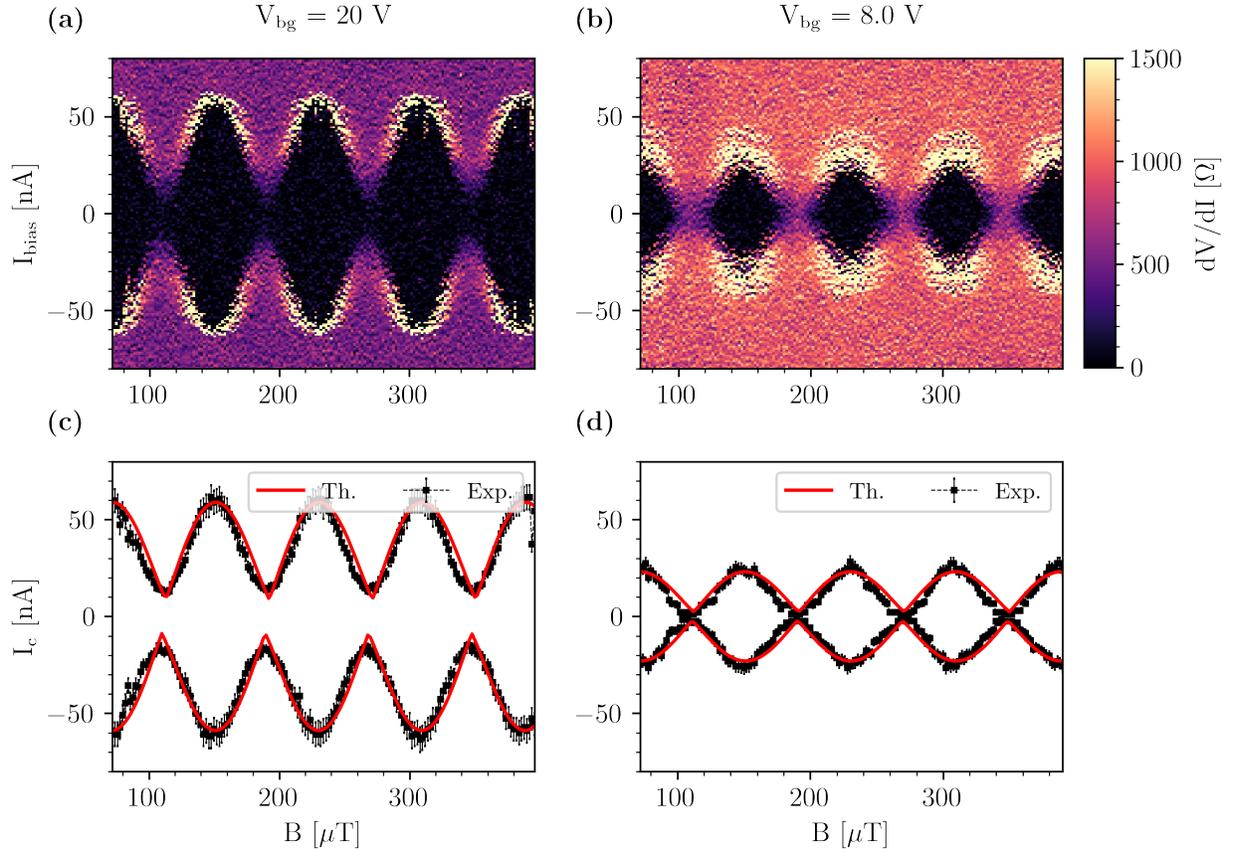

Figure 2: Magnetic field response of the symmetric SQUID. (a), (b) Color maps of the differential resistance $dV/dI$ as a function of current bias and the applied magnetic field. (a) $V_{bg} = 20\,\text{V}$, (b) $V_{bg} = 8\,\text{V}$. The temperature is $T = 350\,\text{mK}$. (c), (d) Comparison of the critical current between experimental data and theoretical model (see text) for the two back-gate configurations presented in (a) and (b), respectively.

the SQUID loop.

A perpendicular magnetic field is applied to the sample to measure the interference pattern, and the corresponding $V - I$ curves are measured while the magnetic field is swept. The results for symmetric SQUID are presented in Fig. 2, which shows the interference pattern for two values of $V_{bg}$. At $V_{bg} = 20\,\text{V}$, shown in Fig. 2(a), the switching current modulates between $60\,\text{nA}$ and $10\,\text{nA}$ in a periodic fashion, typical of a SQUID interferometer. Since the supercurrent is not zero for any magnetic field, only partial and not complete destructive interference is observed at this back-gate voltage. However, when the back-gate voltage is decreased, as shown in Fig. 2(b), the supercurrent modulates completely to zero, and there is full destructive interference. The period of the interference pattern ($\Delta B$) is the same for both voltages on the back-gate. The effective area of the SQUID is obtained as $A_{\text{eff}} = \Phi_0/\Delta B$, yielding $A_{\text{eff}} = 26\,\mu\text{m}^2$, 1.9 times larger than the geometrical area of the loop, $A_{\text{geo}} = 13.6\,\mu\text{m}^2$. This difference can be attributed to the flux-focusing effect of the superconducting strips. Due to the Meissner effect, the strips partially deviate the flux density $\vec{B}$ into the loop, effectively increasing the enclosed flux.[42]

To understand the observed features, we performed numerical simulations. Details are provided in Section S4 of the SI. At low energy, the electronic properties of the InSb nanoflags can be described by a two-band model[43]

$$\mathcal{H}(\vec{k}) = \left(\frac{\hbar^2 \vec{k}^2}{2m^*} - \mu\right)\sigma_0 - \alpha_R k_y \sigma_x + \alpha_R k_x \sigma_y + \mathcal{E}_B \sigma_z, \qquad (5)$$

where $\sigma_\nu$, $\nu \in \{0, x, y, z\}$, are the Pauli matrices acting on the spin degree of freedom. In the above equation, $m^*$ is the effective electron mass in InSb, $\alpha_R$ the Rashba spin-orbit coupling, $\mu$ the chemical potential, and $\mathcal{E}_B = \frac{1}{2}g\mu_B B$ the Zeeman energy. The material parameters are listed in Section S5 of the SI. The chemical potential is directly proportional to the voltage on the back-gate $\mu \propto V_{\text{bg}}$ using a simple capacitive model (see Section S5 of the SI).

The single Josephson junction is described by the Bogoliubov-de Gennes Hamiltonian

$$\mathcal{H}_{\text{BDG}}(x) = \begin{pmatrix} \mathcal{H} & \Delta(x) \\ \Delta^*(x) & -\mathcal{T}\mathcal{H}\mathcal{T}^{-1} \end{pmatrix}, \qquad (6)$$

where $\mathcal{T} = -i\sigma_y \mathcal{K}$ is the operator implementing time-reversal symmetry and

$$\Delta(x) = \Delta\big[\theta(-x) + e^{i\phi}\theta(x - L)\big], \qquad (7)$$

with $\Delta$ the induced superconducting gap and $\theta(x)$ the step function. To compute the Josephson current of the SQUID, we model two JJs in parallel by regularizing Eq. 6 onto a tight-binding model on a square lattice. Here, we add the orbital effects associated with the magnetic field through the Peierls substitution. Then, we employ the recursive Green's function formalism[44,45] to compute the Josephson response (see Section S4 in the SI for details).

In Fig. 2(c)-(d), we show a comparison of the critical currents resulting from the experimental data extracted from Fig. 2(a)-(b), respectively, and the numerical simulations. The simulation parameters can be found in Section S5 of the SI. The theoretical simulations correctly reproduce the closing of the SQUID pattern (complete destructive interference) at low back-gate voltages. This can be explained as follows: at high back-gate voltages, the CPRs are skewed, resulting in a partially destructive interference; at lower back-gate voltages, on the other hand, the junction transparencies decrease with a correspondingly more sinusoidal behavior of the CPRs. This interpretation is supported by the CPR plots of one of the junctions at zero field reported in Fig. 3(a). Here, it can be seen that the CPR at $V_{\rm bg} = 20$V deviates significantly more from a sinusoidal behavior than at $V_{\rm bg} = 8$V. To further corroborate this statement, we performed a Fourier analysis of the CPRs as shown in Fig. 3(b). Here, the ratio of the spectral weights of the $k$-th harmonics with respect to the fundamental tone is plotted on a logarithmic scale. A significant contribution from higher harmonics is present, especially at high back-gate voltages. We stress that this is consistent with a phenomenological relation between the junction transparency and the back-gate voltage derived from the Blonder-Tinkham-Klapwijk (BTK) model,[46] $\tau = 1/(1 + Z^2(V_{\rm bg}))$ (see Sections S4 and S5 of the SI for more details), whose behavior is reported in the inset to Fig. 3(b).

A different phenomenology is observed in the asymmetric device of Fig. 1(b). Figure 4(a) shows the interference pattern at $V_{\rm bg} = 12$ V, for which characteristic SQUID-type interference is present, and the critical current modulates from $I_{\rm c} \sim 100$ nA to $\sim 30$ nA. The periodicity displayed corresponds to an effective area of $A_{\rm eff} = 149\,\mu{\rm m}^2$, 1.26 larger than $A_{\rm geo} = 118\,\mu{\rm m}^2$. The flux-focusing factor of this device is slightly lower than that of the symmetric SQUID, which is expected because the asymmetric SQUID has a larger area. Consequently, the ratio of the area of the Nb strips to the loop area is reduced, and as such, the relative amount of magnetic field screened is also diminished.

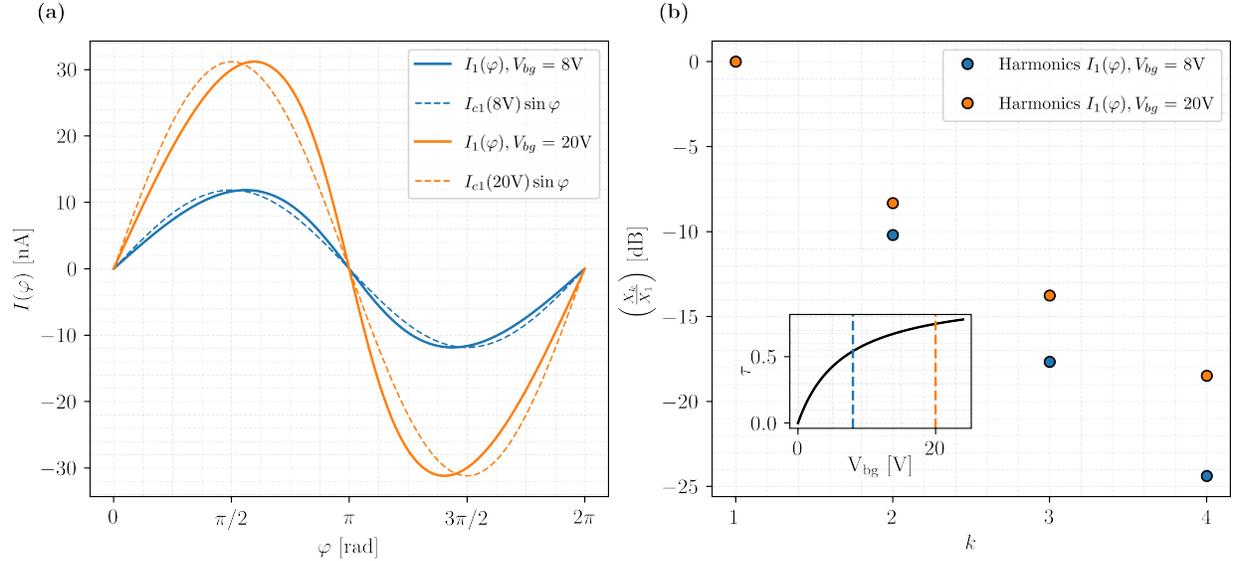

Figure 3: Single junction CPR at zero field. (a) CPR of one of the junctions forming the symmetric SQUID, for $V_{bg} = 20$V (orange) and $V_{bg} = 8$V (blue). Dashed lines correspond to fully sinusoidal CPRs with the same critical current amplitudes for comparison. (b) Ratio of the spectral weights $X_k$ with respect to the first harmonic $X_1$ expressed in dB, for the CPRs in (a). The inset shows a plot of the junction transparency versus back-gate voltage.

When $V_{bg}$ is reduced, as shown in Fig. 4(b) and Fig. 4(c), the modulation amplitude of the interference pattern decreases, completely disappearing at $V_{bg} = 4$ V. This last observation indicates that one of the two Josephson junctions is pinched off and below threshold. Thus, the interferometer has only one arm available for transport, and the supercurrent, not enclosing any magnetic flux, does not show a SQUID-like interference. All these features are well captured by the theoretical model described above, adopted for the asymmetric configuration, as shown in panels (d-f) of Figure 4. Again, both junctions forming the SQUID possess skewed CPRs at high back-gate voltages. It is worth highlighting that the SQUID patterns shown in Fig. 4(a-b) do not show complete destructive interference due to the asymmetry in the geometry of this SQUID device. Section S6 of the SI shows more interference patterns for other back-gate voltages.

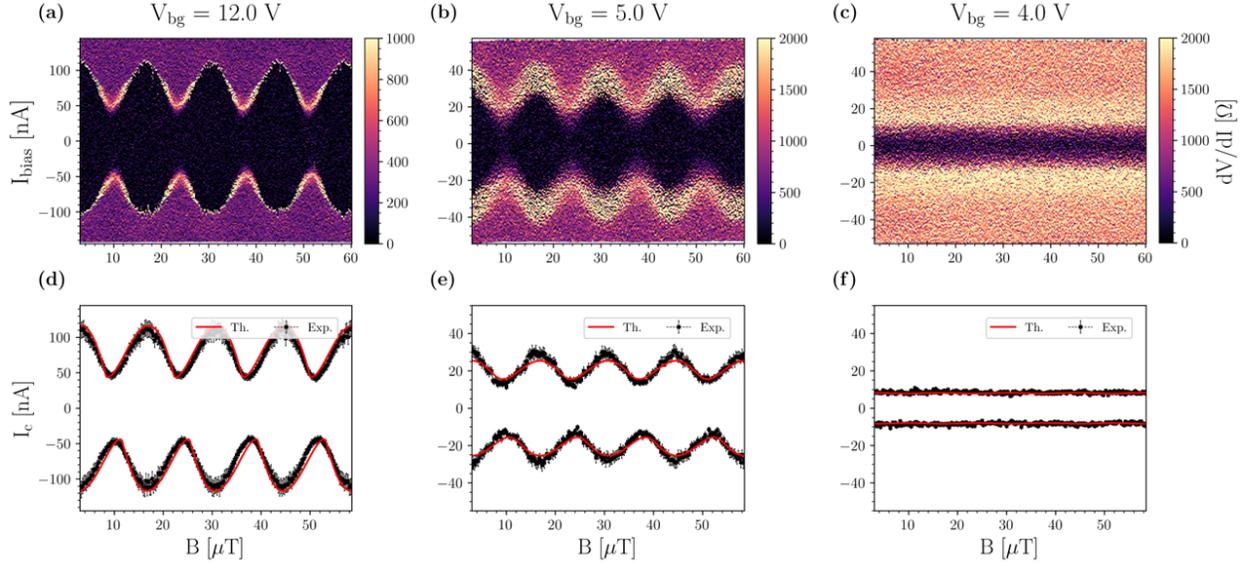

Figure 4: Magnetic field response of the asymmetric SQUID. (a), (b), (c) Color maps of the differential resistance $dV/dI$ as a function of current bias and the applied magnetic field. (a) $V_{bg} = 12\,\text{V}$, (b) $V_{bg} = 5\,\text{V}$, and (c) $V_{bg} = 4\,\text{V}$. $T = 350\,\text{mK}$. (d), (e), (f) Comparison of the critical currents between the SQUID experimental data and the theoretical model for the configurations in (a), (b), and (c), respectively.

Upon closer inspection of Fig. 4(a), it can be seen that there is a difference in critical current when sweeping the current bias up ($I_{c+}$) or down ($|I_{c-}|$). The magnetic fields at which $I_{c+}$ and $|I_{c-}|$ have minima are not the same. This behavior is characteristic of non-reciprocal transport, showing the so-called Josephson diode effect.[28,38] Consequently, at a fixed magnetic field there is a range of current bias values where the SQUID is superconducting in one current bias direction and dissipative in the other. This feature is related to the asymmetric configuration, and can be quantified by the rectification coefficient $\eta = (I_{c+} - |I_{c-}|)/(I_{c+} + |I_{c-}|)$ which reaches 6% here. The rectification coefficient is also found to be gate-tunable and goes to zero for $V_{bg} = 4\,\text{V}$. This feature connects the Josephson diode effect to the semiconducting nanoflags and excludes other sources of superconducting diode, such as vortices or superconductor screening currents.[47–49] We note that consistent results were recently found in InSb nanosheet interferometers.[38] Additional details on the Josephson diode effect for this SQUID device can be found in Section S7 of the SI, while Section S8 shows Fraunhofer-like interference patterns in a wider range of magnetic field

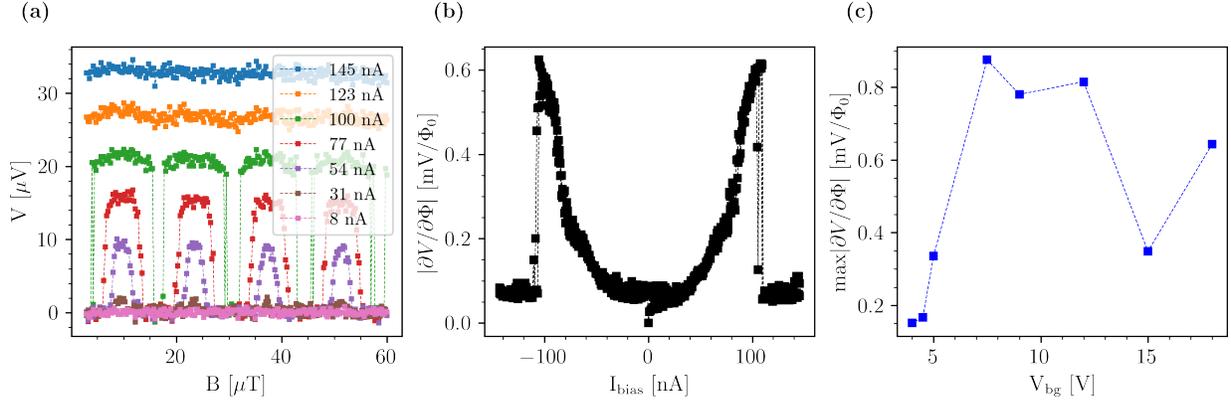

Figure 5: Transfer function characteristics of the asymmetric SQUID. $T = 350\,\text{mK}, V_{bg} = 18\,\text{V}$. (a) $V - B$ characteristics for different values of current bias. (b) Voltage responsivity as a function of current bias. (c) Maximum responsivity as a function of back-gate voltage.

values.

SQUIDs in the non-hysteretic regime are commonly employed as flux-to-voltage transducers to measure small variations in the magnetic field. To test the performance of these SQUIDs as a magnetometer, they are current biased in the dissipative state, at a working point where the $V - B$ response is optimal. We have obtained very similar results for the two geometries. Here we discuss only the asymmetric SQUID. Details on symmetric SQUID can be found in Section S9 of the SI. In Fig. 5(a) we report the $V - B$ curves of the asymmetric SQUID for different values of current bias at $V_{\text{bg}} = 18\text{V}$. For a current bias of 100 nA, a modulation amplitude of about $20\,\mu\text{V}$ is obtained. We use the voltage responsivity to characterize the voltage response of the device. This quantity, which is a standard figure of merit of a magnetometer, is defined as

$$V_\Phi = \left.\frac{\partial V}{\partial \Phi}\right|_{I_{\text{bias}}} \qquad (8)$$

and can be used to relate the voltage noise amplitude $S_V^{1/2}$ to the magnetic flux noise amplitude $S_\Phi^{1/2}$, using $S_V^{1/2} = V_\Phi \cdot S_\Phi^{1/2}$.[2] These quantities are known to be a function of frequency. In particular, the voltage noise amplitude tends to increase as the frequency is decreased, and the $1/f$ noise contribution dominates the white noise spectrum typical of thermal noise.[3,50] DC transport measurements reported here can provide an estimate of the magnetic flux noise at low frequency, $f < 10\,\text{Hz}$, limited by the multimeter integration time.

In Fig. 5(b) we show the measured voltage responsivity as a function of the current bias for $V_{\text{bg}} = 18\,\text{V}$. A maximum of $V_\Phi = 0.6\,\text{mV}/\Phi_0$ is obtained. This value can be further improved by tuning the back-gate, as shown in Fig. 5(c). A maximum of $0.9\,\text{mV}/\Phi_0$ is found for $V_{\text{bg}} = 7.5\,\text{V}$. To convert this quantity to magnetic flux noise, an evaluation of the voltage noise of the experimental setup is needed. In our setup at $T = 350\,\text{mK}$, the main source of noise is the input noise of the room-temperature preamplifier, $S_{V,\text{preamp}}^{1/2} = 4\,\text{nV}/\sqrt{\text{Hz}}$, which finally gives $S_\Phi^{1/2} = 4.4 \times 10^{-6}\,\Phi_0/\sqrt{\text{Hz}}$. This value is consistent with the performance of commercially available superconducting magnetometers[51] and is comparable to those documented in the literature, encompassing both all-metallic configurations[6,52–56] and InAs-based interferometers.[57,58] This comparison underscores the superior quality of these devices. With further optimization, the sensitivity of SQUIDs fabricated from InSb nanoflag Josephson junctions has the potential to be enhanced for application in magnetometry.

In this study, superconducting quantum interference devices realized using InSb nanoflag Josephson junctions have been thoroughly investigated. Two distinct geometries were characterized, in which the dual junctions forming the SQUID are arranged in either symmetric or asymmetric configurations. An appropriate theoretical framework adequately accounts for all observed phenomena. Specifically, it was demonstrated that the transparency of the junction can be modulated by a back-gate voltage, leading to partial destructive interference in the symmetric configuration at elevated voltages and complete destructive interference at reduced voltages. Furthermore, the higher harmonic content of these junctions, which yields skewed current-phase relations (CPRs), has been quantified. The asymmetric configuration exhibits nonreciprocal supercurrent transport with a diode efficiency of $\sim 6\%$, which is consistent with a recent report.[38] This observation further corroborates the higher harmonic content and superior quality of these junctions. Additionally, the SQUID performance has been evaluated, particularly their low-frequency flux-to-voltage response, highlighting their potential applicability as nanoscale magnetometers.

# Acknowledgement

The authors acknowledge support from project PRIN2022 2022-PH852L(PE3) TopoFlags-"Non-reciprocal supercurrent and topological transition in hybrid Nb-InSb nanoflags" funded by the European community-Next Generation EU within the program "PNRR Missione 4-Componente 2-Investimento 1.1 Fondo per il Programma Nazionale di Ricerca e Progetti di Rilevante Interesse Nazionale (PRIN)" and by PNRR MUR Project No. PE0000023-NQSTI. F.G. acknowledges the EU's Horizon 2020 Research and Innovation Framework Programme under Grants No. 964398 (SUPERGATE) and No. 101057977 (SPECTRUM) for partial financial support.

# Supporting Information Available

The Supporting Information is available free of charge.

- Fabrication methods, Measurement methods, Inductance Estimation, Modelling of a single Josephson junction, Simulation of the SQUID setup, Additional Data, Josephson Diode Effect in the asymmetric SQUID, Fraunhofer-like interference, and Magnetometer response for the symmetric SQUID.

# Supporting Information for "Superconducting Quantum Interference Devices based on InSb nanoflags Josepshon junctions"

Andrea Chieppa,[1] Gaurav Shukla,[1] Simone Traverso,[2,3] Giada Bucci,[1] Valentina Zannier,[1] Samuele Fracassi,[2,3] Niccolo Traverso Ziani,[2,3] Maura Sassetti,[2,3] Matteo Carrega,[3] Fabio Beltram,[1] Francesco Giazotto,[1] Lucia Sorba,[1] and Stefan Heun[1,*]

[1] *NEST, Istituto Nanoscienze-CNR and Scuola Normale Superiore, Piazza San Silvestro 12, 56127 Pisa, Italy*

[2] *Dipartimento di Fisica, Università di Genova, Via Dodecaneso 33, 16146 Genova, Italy*

[3] *CNR-SPIN, Via Dodecaneso 33, 16146 Genova, Italy*

(Dated: April 24, 2025)

## S1. FABRICATION METHODS

The fabrication process of the SQUIDs begins with the as-grown sample of InSb nanoflags, which are attached to the Indium Phosphide stems [1]. The initial step involves transferring the free-standing structures onto a highly conductive p-type Si(100) substrate, serving as a global back-gate. A 285 nm thick SiO$_2$ layer covers the Si substrate, acting as a dielectric.

A standard electron beam lithography (EBL) technique is used for fabricating the Nb contacts. First, 270 nm of AR 679.04 resist is spin-coated at 4000 rpm for 1 minute, followed by baking the sample at 170 °C for 90 s. EBL is performed at 20 kV accelerating voltage of electrons, 10 µm aperture, ≈ 33 pA current, and 290 µC cm$^{-2}$ dose using a 200 × 200 µm$^2$ write field. After EBL, the pattern is developed in AR 600-56 for 1 minute, followed by rinsing in IPA for 30 seconds prior to drying with N$_2$ flux. The developed pattern is then exposed to O$_2$ plasma (15 W for 75 seconds) for descum to remove any residual resist in the pattern.

Prior to sputtering Niobium on the EBL-patterned sample, to achieve ohmic contacts [2], the exposed area of the InSb nanoflags is passivated by immersing the sample in (NH$_4$)$_2$S$_x$ (290mM(NH$_4$)$_2$S and 330mM S in deionized water) at 45 °C for 60 seconds, followed by cleaning in deionized water for 30 seconds prior to drying with N$_2$ flux.

Sputtering of 180 nm of Niobium is performed at 150 W for 240 seconds (at a rate of 7.5 Å/s) at a base pressure of $8 \times 10^{-8}$ mbar and a working pressure of $5 \times 10^{-3}$ mbar in the presence of Ar. The sputtered sample is then kept in acetone overnight for the lift-off process, followed by cleaning in IPA for 30 seconds.

After completing the fabrication process, the sample chip is glued to a dual-in-line chip carrier using highly conductive silver paste, which enables the operation of the back-gate. The individual SQUIDs are then connected to the chip carrier by Aluminum wire bonding.

The two geometries introduced in the main text are displayed in the scanning electron micrographs in Fig. S1. The geometrical parameters are given in Table S1.

* stefan.heun@nano.cnr.it

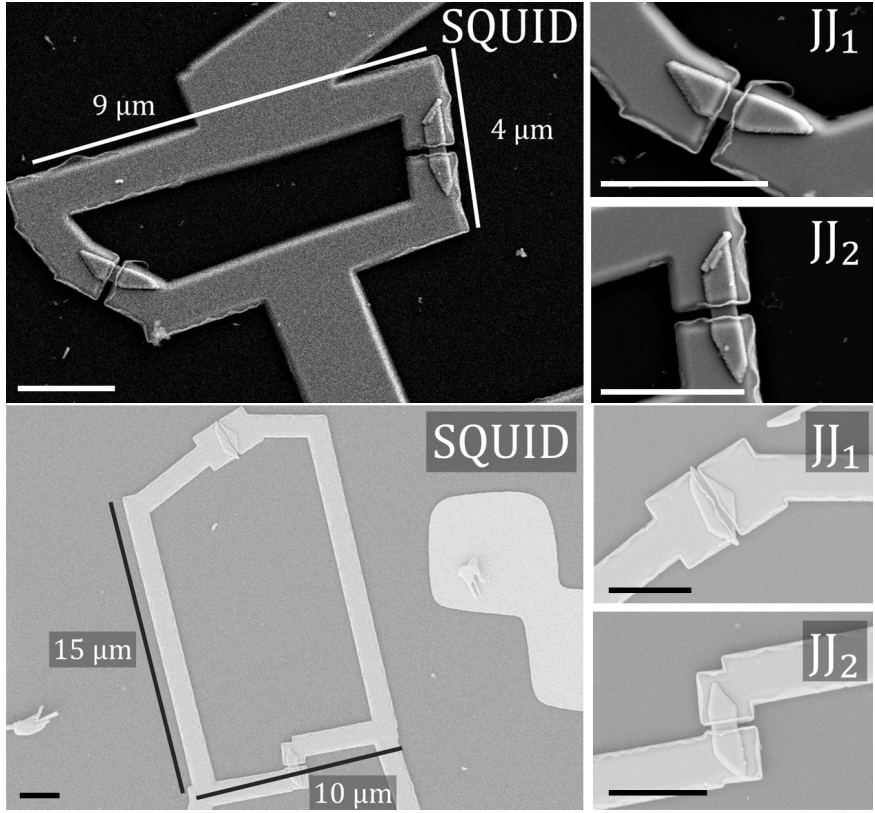

FIG. S1. SQUID in a (a) symmetric geometry and (b) asymmetric geometry. A zoom on both Josephson junctions is displayed in the right panels. Scale-bars in the lower left corner of each panel indicate 2 µm.

|  | L1 | L2 | W1 | W2 | $A_{JJ1}$ | $A_{JJ2}$ | $A_{loop}$ |
|---|---|---|---|---|---|---|---|
| Symmetric SQUID | 200 | 200 | 380 | 380 | 0.11 | 0.11 | 13.6 |
| Asymmetric SQUID | 180 | 190 | 1700 | 530 | 0.44 | 0.14 | 118 |

TABLE S1. Geometrical parameters of the devices presented in this work. Subscript $i$ refers to the Josephson junction $i$, shown in Fig. S1 (a) and (b). Lengths in nm, areas in µm$^2$. For the area of the junctions, $A_{JJi}$, twice the London penetration depth of Niobium has been included ($\lambda_L = 43$ nm [3]).

## S2. MEASUREMENT METHODS

Magneto-transport measurements have been performed in an Ice-Oxford dry cryostat, with a continuos-mode base temperature of $T = 350\,\text{mK}$. The conductance measurements shown in Fig. 1(b) and (c) of the main text are performed at $T = 2\,\text{K}$ with an AC current bias setup, using two SR-830 Lock-In Amplifiers. The back-gate and the solenoid superconducting magnet are powered with two Keithley model 2400B Souce Measure Unit. The voltage-current characteristics were acquired in current bias feeding a current with a Yokogawa GS200 generator and employing two Stanford Research preamplifiers, model SR560 and SR570, measuring the amplified signals with two Agilent 34401A multimeters.

## S3. INDUCTANCE ESTIMATION

The magnetic flux enclosed in the SQUID loop is the sum of an external applied flux and an induced flux. The latter is due to the self-inductance $L$, which is the sum of the geometrical contribution $L_{geo}$ and the kinetic contribution $L_{kin}$, i.e., $L = L_{geo} + L_{kin}$. When the SQUID supports supercurrent transport, a finite value of $L$ shifts the working point in the magnetic flux by $LI_{circ}$, where $I_{circ}$ is the supercurrent circulating in the loop. To understand the relevance of this effect, the inductance parameter $\beta_L = 2\pi \frac{LI_c}{\Phi_0}$ is a useful quantity. Usually, when $\beta_L$ ranges from $10^{-1}$ to $1$, the interpretation of an interference pattern must explicitly take into account a finite $L$, and precise calculations of it are needed. This often goes through numerical simulations and finite element methods. However, a prior estimate of $L$ is needed to understand if it produces any significant effect.

This is done calculating the geometrical contribution, $L_{geo}$, under the assumption of a rectangular loop (sides $l, w$) with a circular cross section (radius $r_w$), that let us use [4]:

$$L_{geo} = \frac{\mu_0}{\pi} \left[ -(l-r_w)\sinh^{-1}\frac{l-r_w}{w-r_w} - (w-r_w)\sinh^{-1}\frac{w-r_w}{l-r_w} + (l-r_w)\sinh^{-1}\frac{l-r_w}{r_w} + \right.$$
$$+ (w-r_w)\sinh^{-1}\frac{w-r_w}{r_w} + r_w\sinh^{-1}\frac{r_w}{w-r_w} + r_w\sinh^{-1}\frac{r_w}{l-r_w} + 2\sqrt{(l-r_w)^2+(w-r_w)^2}$$
$$\left. -2\sqrt{(w-r_w)^2+(r_w)^2} - 2\sqrt{(l-r_w)^2+(r_w)^2} - 2r_w\ln\left(1+\sqrt{2}+2\sqrt{2r_w}\right) \right] \quad (1)$$

For instance, using $l = 9\,\mu\text{m}$, $w = 4\,\mu\text{m}$, and $r_w = 0.36\,\mu\text{m}$ for the symmetric SQUID, leads to $L_{geo} = 9.7\,\text{pH}$.

The kinetic contribution $L_{kin}$ arises due to the kinetic energy of Cooper pairs, and for a superconducting wire of length $l$, width $w$, and thickness $t$, can be estimated with the following formula [5, 6]:

$$L_{kin} = \frac{\mu_0 \lambda_L^2 l}{wt} \quad (2)$$

In practical situations, this contribution is relevant only when the superconducting strips have thickness comparable to or smaller than the London penetration depth $\lambda_L$. In our case, the strips have thickness $t = 150\,\text{nm}$, larger than $\lambda_L = 43\,\text{nm}$ [3], and an explicit calculation for the symmetric SQUID gives $L_{kin} = 0.3\,\text{pH}$.

For this device, a total inductance of $10\,\text{pH}$ is found, which together with the upper limit on $I_c \sim 100\,\text{nA}$ ultimately leads to a $\beta_L$ of the order of $10^{-3}$. The same procedure for the asymmetric device gives also a $\beta_L$ of the order of $10^{-3}$. Hence, we conclude that corrections to the magnetic flux owing to the self-inductances of the loops can be considered negligible.

## S4. MODELLING OF A SINGLE JOSEPHSON JUNCTION

We consider a planar Josephson junction in the $x-y$ plane and extending along the $x$-direction, with the left and right leads obtained by proximitizing the InSb nanoflag with a conventional $s$-wave superconductor. We denote by $L$ the junction length, and by $W$ its width. Furthermore, we denote by $(x_0, y_0)$ the coordinates of the site at the bottom left corner of the scattering region.

First, to obtain a tight-binding Hamiltonian for the InSb nanoflag in the normal region, we regularize the continuum Hamiltonian of the main text [Eq. (5)] on a square lattice with lattice constant $a$. The resulting $\vec{k}$-space Bloch Hamiltonian is given by

$$\mathcal{H}(\vec{k}) = \{(4t-\mu) - 2t[\cos(k_x) + \cos(k_y)]\}\sigma_0 - 2\mathcal{E}_R \sin(k_y)\sigma_x + 2\mathcal{E}_R \sin(k_x)\sigma_y + \mathcal{E}_B \sigma_z, \quad (3)$$

where $\sigma_0$ is the $2\times 2$ identity matrix, and $\sigma_i$, $i=x,y,z$ are the Pauli matrices acting on the spin degree of freedom. Here the Bloch momentum is adimensional, $t = \frac{\hbar^2}{2m^* a^2}$ parametrizes the first neighbor hopping, and $m^*$ is the electron effective mass. Moreover, $\mathcal{E}_B$ and $\mathcal{E}_R$ are

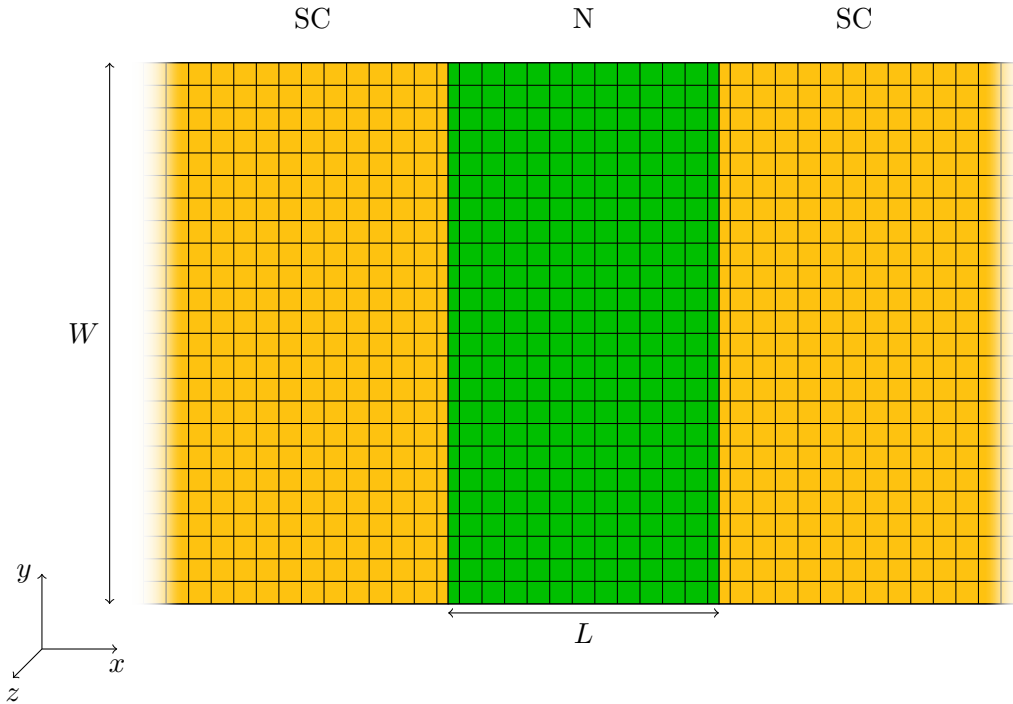

FIG. S2. Scheme of a planar Josephson junction, overlaid the discretized square lattice and located in the $x-y$ plane. The normal region is colored in green, while the superconducting leads are in yellow.

energy scales associated to the magnetic field and to the Rashba spin-orbit coupling. They are defined as

$$\mathcal{E}_B = \frac{1}{2}g\mu_B B, \tag{4}$$

$$\mathcal{E}_R = \frac{\alpha_R}{2a}. \tag{5}$$

with $g$ the giromagnetic factor, $\mu_B$ the Bohr magneton, and $\alpha_R$ the Rashba coupling.

We now write down the real space tight-binding Hamiltonian for the junction. Starting from the Fourier transform of Eq. (3), we add magnetic field related orbital effects and random disorder to the normal region, $s$-wave superconducting pairing to the proximitized leads, and potential barriers at the interfaces. We denote by $\tilde{L} = L/a$ and $\tilde{W} = W/a$ the number of sites along the horizontal and vertical dimension of the scattering region, and we associate the lattice indexes $(\ell, j) = (0, 0)$ to the bottom-left site at coordinate $(x_0, y_0)$. Then, the real space Hamiltonian for the junction can be written as $H = H_0 + H_{\rm SC}$, where

$$H_0 = \sum_{\ell,j} \Psi^\dagger_{\ell,j} H_{\rm o.s.}(\ell,j) \Psi_{\ell,j} + (\Psi^\dagger_{\ell,j+1} V_y \Psi_{\ell,j} + {\rm h.c.}) + (\Psi^\dagger_{\ell+1,j} V_x(\ell,j) \Psi_{\ell,j} + {\rm h.c.}), \tag{6}$$

$$H_{\rm SC} = \Delta_\ell c^\dagger_{\uparrow \ell,j} c^\dagger_{\downarrow \ell,j} + {\rm h.c.}, \tag{7}$$

with the spinor $\Psi^T_{\ell,j} = (c_{\uparrow \ell,j}, c_{\downarrow \ell,j})$ collecting the operators $c_{\uparrow \ell,j}$ and $c_{\downarrow \ell,j}$, that in turn destroy an electron with spin up and spin down, respectively, at the site indexed by $(\ell, j)$.

Let us break down the various terms appearing in the Hamiltonian, starting from $H_{\rm SC}$. This accounts for the induced $s$-wave superconductivity in the leads, and $\Delta_\ell$ is given by

$$\Delta_\ell = \begin{cases} \Delta & \ell < 0 \\ 0 & 0 \leq \ell < \tilde{L} \\ \Delta e^{i\phi} & \ell > \tilde{L} \end{cases}, \tag{8}$$

with $\Delta$ the induced gap amplitude, and $\phi$ the superconducting phase difference between the two leads.

Regarding $H_0$, instead, the matrices $H_{\rm o.s.}$, $V_x$, and $V_y$ have the explicit expressions

$$H_{\rm o.s.}(\ell,j) = (4t - \mu + \omega_{\ell,j} + \mathcal{U}_\ell)\sigma_0 + \frac{1}{2}g\mu_B B_\ell \sigma_z, \tag{9}$$

$$V_x(\ell,j) = [-t\sigma_0 + i\mathcal{E}_R \sigma_y] e^{-i\frac{e}{\hbar} B_\ell(ja - y_0)a}, \tag{10}$$

$$V_y = -t\sigma_0 - i\mathcal{E}_R \sigma_x. \tag{11}$$

The magnetic field spatial dependence is given by

$$B_\ell = \begin{cases} 0 & \ell < 0 \\ B & 0 \leq \ell < \tilde{L} \\ 0 & \ell > \tilde{L} \end{cases}, \qquad (12)$$

that is, the magnetic field is assumed to be uniform in the normal region and zero in the leads. The orbital effects are included via the Peierls phase in $V_x$, which results from the choice of the Landau gauge $\vec{A} = (-By, 0, 0)$ for the vector potential.

The on-site energies $\omega_{\ell,j}$ are added to account for possible substrate-induced disorder in the normal region ($\omega_{\ell,j} = 0$ for $\ell < 0$ or $\ell \geq \tilde{L}$), and are randomly extracted from the interval $[-V_{\text{dis}}/2, V_{\text{dis}}/2]$ ($V_{\text{dis}} > 0$) with an uniform distribution.

Finally, the term $\mathcal{U}_\ell = \mathcal{U}(\delta_{0,\ell} + \delta_{\tilde{L}-1,\ell})$ models two potential barriers, placed on the first and last column of sites of the normal region, respectively. The barriers are added to tune the transparency at the NS interface, which according to the Blonder-Tinkham-Klapwijk (BTK) model [7] is defined as

$$\tau = \frac{1}{1+Z^2}, \qquad (13)$$

with $Z = m^*\mathcal{U}a/(\hbar^2 k_F)$ and $k_F$ the Fermi momentum. If we assume a parabolic dispersion of the bands (which is indeed the case at low energy) and $\mu = \hbar^2 k_F^2/(2m^*)$, then we have $Z = \sqrt{\frac{m^*}{2\mu\hbar^2}}\mathcal{U}a$.

Given the tight-binding Hamiltonian of both the leads and the scattering region in Eq. (6), the equilibrium Josephson supercurrent is computed through the recursive Green's function approach [8, 9]. Concerning the surface Green's functions of the semi-infinite uncoupled leads, these are computed via the infinite recursive Green's function method [10].

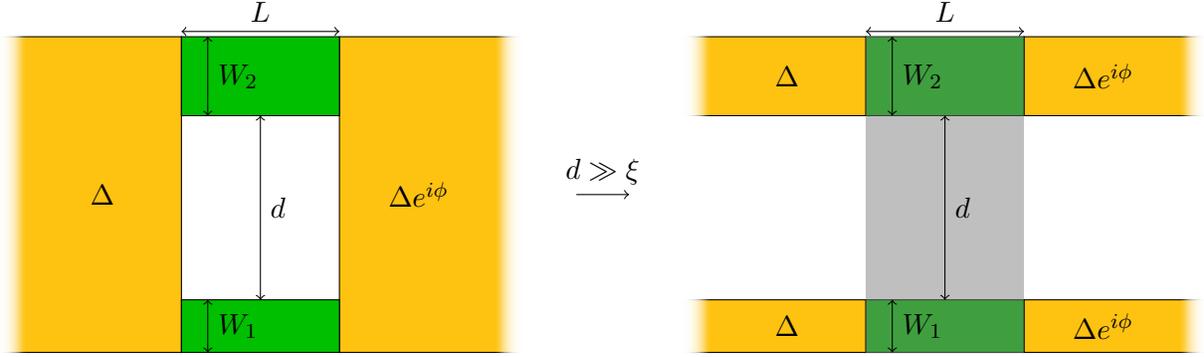

FIG. S3. Scheme of the SQUID setup considered for the simulations. The SQUID geometry on the left, with the superconducting leads covering the left and right ends of the two nanoflags, can be simulated as two independent junctions whenever the separation between the flags is much larger than the superconducting coherence length. This effective setup is shown on the right. The transparent gray overlay covering the normal regions of the flags and the region between them indicates the region where the magnetic field is present.

## S5.  SIMULATION OF THE SQUID SETUP

We model the SQUID setup as shown in Fig. S3. Here, two junctions of width $W_1$ and $W_2$ are parallel to each other and have the same length $L$. We denote the separation between the two junctions by $d$. Although the superconducting loop geometry is markedly different with respect to that of the actual experimental devices, this is not an issue. Indeed, the response of a SQUID should not change due to deformations of the superconducting loop, as long as the magnetic flux piercing through it is kept constant.

Furthermore, to numerically compute the supercurrent in the SQUID configuration we resort to the following approximation: since the two junctions forming the SQUID are separated by a distance $d$ much larger than the coherence length of the superconductor ($d \gg \xi$), they can be regarded as effectively decoupled (see the right part of Fig. S3). Thus, we actually compute the supercurrent of two independent JJs having same length $L$, widths $W_1$ and $W_2$, and, most importantly, presenting the same superconducting order parameter on the left ($\Delta$) and right ($\Delta e^{i\phi}$) sides, respectively. Crucially, the spatial separation between the junctions (and consequently the field flux in the loop area) is fully accounted for by the $y$-dependence in the Peierls substitution, correctly yielding the expected interference pattern. In fact, if we fix $y = 0$ at the bottom of the lower junction, then the bottom of the upper one sits at $y = W_1 + d$. These different offsets enter as different values of $y_0$ in Eq. (10) for the two junctions. Under these assumptions, the area of the SQUID pierced by the magnetic flux is defined as $A_{\text{eff}}^{\text{th}} = L\left(d + \frac{W_1+W_2}{2}\right)$. In the simulations, we fix $d$ such that that $A_{\text{eff}}^{\text{th}}$ matches the $A_{\text{eff}}$ extracted from the experimental data on the SQUID pattern, so as to correctly reproduce the SQUID periodicity.

| Parameter | Symmetric SQUID | Asymmetric SQUID |
|---|---|---|
| $L$ | 200nm | 180nm |
| $W_1$ | 380nm | 530nm |
| $W_2$ | 380nm | 1700nm |
| $\Delta$ | 280μeV | 300μeV |
| $\mathcal{U}_1$ | 55meV | 40meV |
| $\mathcal{U}_2$ | 57meV | 58meV |
| $V_{\text{dis}}$ | 10meV | 10meV |
| $C$ | 4.11meV V$^{-1}$ | 7.8meV V$^{-1}$ |
| $V_{\text{th}_1}$ | 0V | 3.8V |
| $V_{\text{th}_2}$ | 0V | 2.56V |

TABLE S2. Numerical values of the model parameters used in the simulations for the InSb nanoflag. The subscripts $j = 1, 2$ indicate the normal region of the two junctions, respectively.

Concerning the chemical potential appearing in the theoretical model, this is assumed to be tuned by the back-gate potential according to the following phenomenological relation

$$\mu = C(V_{\text{bg}} - V_{\text{th}}), \tag{14}$$

with the threshold voltage which can be different for the two normal regions and is set to zero in the superconducting leads.

Table S2 lists the parameters used for the simulation of the symmetric and asymmetric SQUID. Although the parameters have been fine-tuned to better fit the experimental data, it should be underlined that they all possess the correct order of magnitude and that small variations of the reported values do not significantly alter the correspondence with the experimental data.

## S6. ADDITIONAL DATA

Supplementary interference patterns for different back-gate voltages for the symmetric SQUID are shown in Fig. S4. No loss of interference is observed in lowering the back-gate voltage, meaning that in both arms the supercurrent pinches off at the same $V_{bg}$, which is $\sim 4.0\,\text{V}$, consistent with the symmetric design and with the single-threshold picture of the normal-state characterisation.

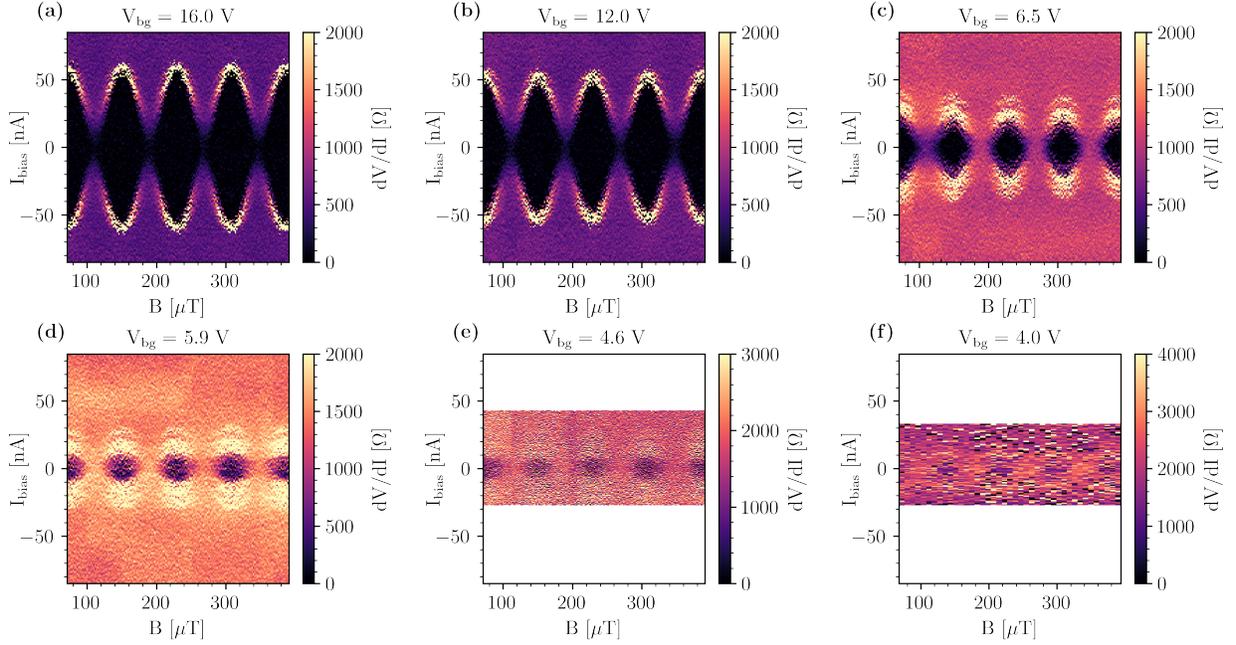

FIG. S4. Symmetric SQUID interference patterns for different back-gate voltages.

The asymmetric SQUID patterns for various $V_{\text{bg}}$ values are shown in Fig. S5. An asymmetry in the critical currents, $I_{c1} \neq I_{c2}$, is observed across all the explored $V_{\text{bg}}$ range. At approximately $V_{\text{bg}} \sim 4.0\,\text{V}$, one of the junctions pinches off and does not support supercurrent transport, leading to the absence of observable SQUID interference. As the chemical potential is brought by the back-gate into the band-gap of the semiconductor, the supercurrent transport of the Josephson junction transits from a SNS-junction regime to an exponentially suppressed tunneling regime, where $I_c \sim 0\,\text{nA}$ given the electrode spacing of $L \approx 200\,\text{nm}$.

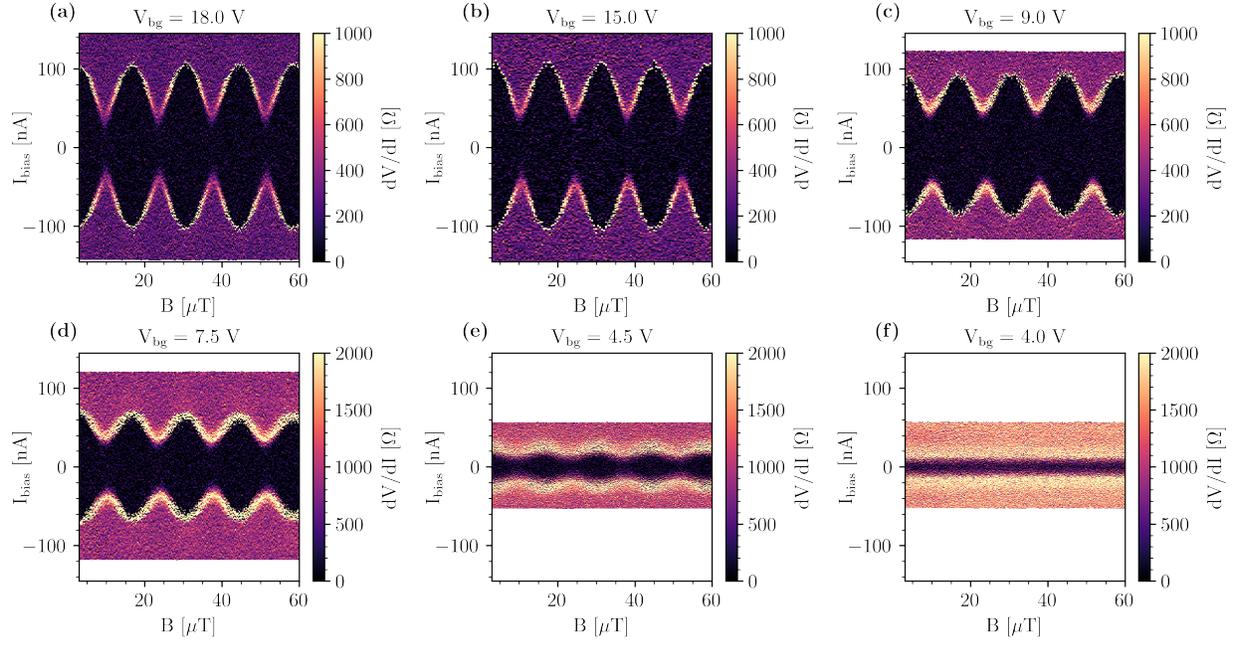

FIG. S5. Asymmetric SQUID interference patterns for different back-gate voltages.

## S7. JOSEPHSON DIODE EFFECT IN THE ASYMMETRIC SQUID

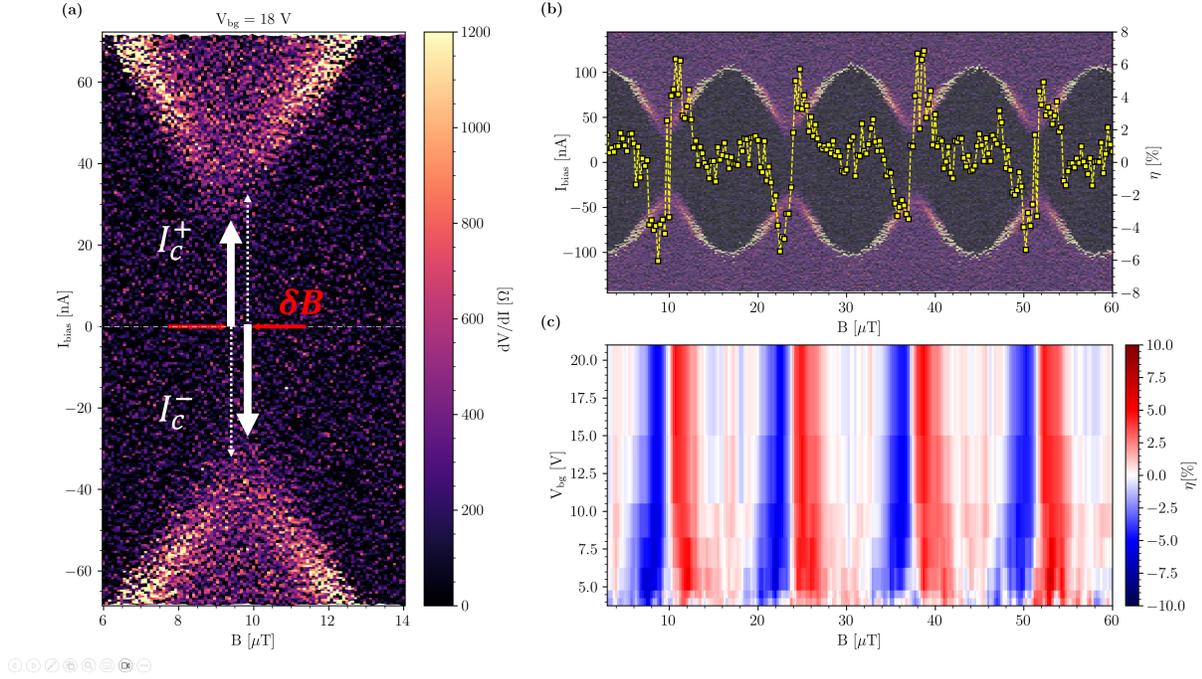

FIG. S6. (a) Zoom near the minima in the interference pattern of the asymmetric SQUID. $T = 350\,\mathrm{mK}, V_{bg} = 18\,\mathrm{V}$. (b) The rectification coefficient $\eta$ (yellow squares) superimposed on the interference pattern, highlighting its periodicity with the applied magnetic field. (c) Colour-map of $\eta$ against the back-gate voltage.

During the measurement of the $V - I$ traces, the current bias was swept both in the positive and the negative direction, allowing to acquire both transitions related to $I_{sw,+}$ and $I_{sw,-}$. In the resulting SQUID interference patterns, it is found that the minima in magnetic field of these two quantities are slightly shifted, indicating the presence of Josephson diode effect (JDE). Fig. S6(a) shows a zoom around the minima in the interference pattern presented in Fig. 4(a) of the main text. In this case, the minima are shifted by $\delta B \sim 1\,\mathrm{\mu T}$, corresponding to $\sim 0.07\,\Phi_0$. To quantify the amount of diode effect, is common to use the rectification coefficient $\eta$:

$$\eta = \frac{I_{c+} - |I_{c-}|}{I_{c+} + |I_{c-}|} \tag{15}$$

This is plotted against the magnetic field in Fig. S6(b), where for clarity the corresponding interference pattern is represented in the background. To improve the signal-to-noise ratio, averaging between switching and retrapping current is performed, since no notable difference is present between those two quantities, and JDE is observed in both retrapping and switching currents. An oscillatory behaviour in $\eta$ with the same periodicity as the SQUID pattern is present, and $\eta$ can be tuned from $\sim 6\%$ to $\sim -6\%$. The magnetic fields for which $\eta$ changes sign are those for which the critical current is almost at its maximum and minimum value. This is particularly evident in Fig. S6(c), where a colour-map illustrates both the magnetic field dynamics and the gate-tunability of $\eta$. The rectification coefficient reaches zero for $V_{\text{bg}} = 4\,\text{V}$. The fact that the back-gate does not reverse the sign of $\eta$ indicates that the Josephson junction with the higher critical current is the same for every $V_{\text{bg}}$ explored. In fact, in SQUIDs where the two $I_{ci}$ vs $V_{\text{bg}}$ cross (Ref. [11]) the JDE pattern reports a sign-reversal at the back-gate value for which inversion symmetry is restored. Our observations are consistent with theoretical predictions expected for SQUIDs: three conditions should be present at the same time for a SQUID to display JDE [12]:

1. The external flux must not be equal to a an integer multiple of half the flux quantum: $\Phi \neq n\frac{\Phi_0}{2}$;

2. The transmission of the Josephson junctions (transparency of the interface) must not be equal;

3. At least one Josephson junction needs to be highly transmitting to have a sizable higher harmonic content in the CPR.

Our results thus provide support that InSb nanoflag-based Josephson junctions are highly transmissive and present higher harmonic content in the CPR.

## S8. FRAUNHOFER-LIKE INTERFERENCE

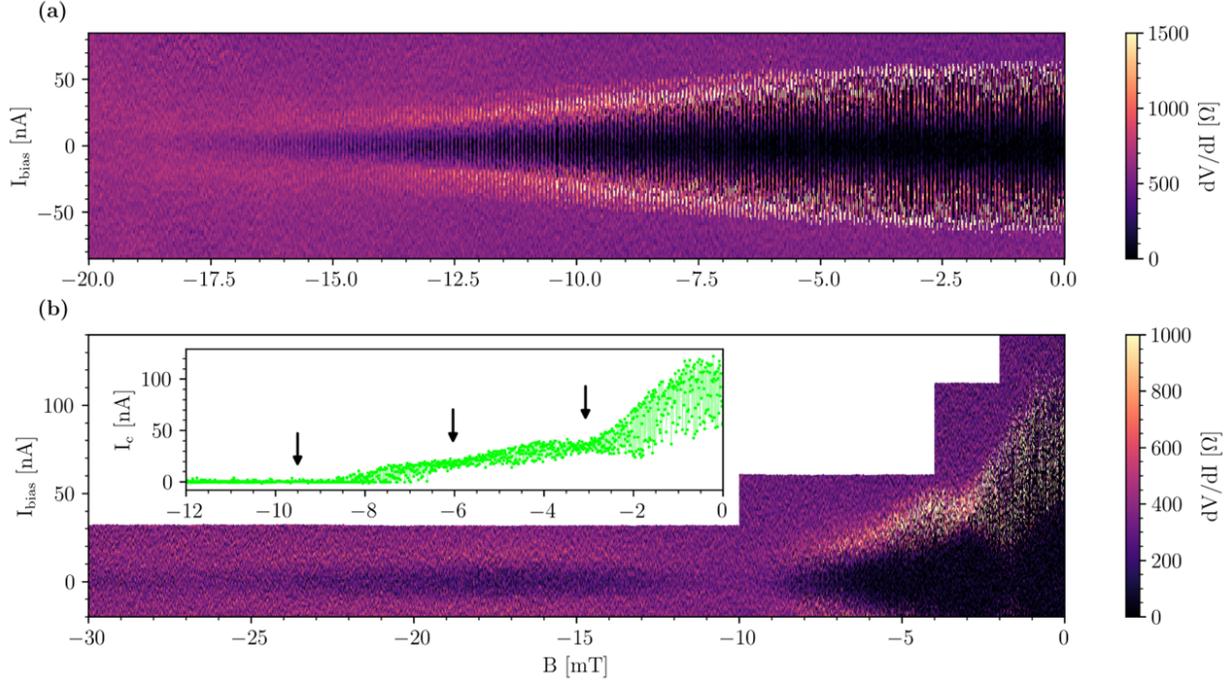

FIG. S7. Fraunhofer patterns of the SQUIDs. (a) Symmetric device, $V_{bg} = 20$ V. (b) Asymmetric device, $V_{bg} = 15$ V. Inset: Critical currents ($I_c$) as a function of magnetic field. SQUID modulation of $I_c$ is locally suppressed at the magnetic field values indicated by black arrows. $T = 350$ mK.

When the magnetic flux over one or both individual junctions, $\Phi_{JJ,i}$, becomes of the same order of magnitude as the superconducting flux quantum, the effects of the SQUID and the Fraunhofer-like interference of the single junctions superimpose, and a low-frequency Fraunhofer envelope modulates the high-frequency SQUID interference pattern. Figure S7 explores the Fraunhofer-like regime of SQUIDs for both geometries, at high back-gate voltage (20 V and 15 V for the symmetric and asymmetric case, respectively). The two individual junctions of the symmetric device modulate the SQUID oscillations with a Gaussian envelope, as shown in Fig. S7(a). For InSb nanoflags, this kind of supercurrent modulation has already been reported for junctions in the narrow configuration [13–15]. In previous works, the absence of pronounced side lobes was attributed to a limited $W/L$ ratio of junctions [15–18]. For the symmetric SQUID, both junctions have $W/L \sim 2$, slightly lower than the previous devices, further supporting this picture. From Fig. S7(a), it is found that at a magnetic field of $B_0 = 18.5$ mT the SQUID interference is no more visible. Along the Fraunhofer pattern, no beating of the modulation amplitude is visible. Hence, both junctions have the same area, consistent with the symmetric geometry. The effective area extracted from this measurement is $0.11\,\mu m^2$, in agreement with the geometrical area of $(L+2\lambda_L) \times W = 0.11\,\mu m^2$ measured from SEM images (with $\lambda_L$ the London penetration depth).

The same measurement for the asymmetric SQUID is displayed in Fig. S7(b). In this case, the Josephson junctions have different areas $(A_{JJ,1}/A_{JJ,2} \approx 3.1)$, resulting in a visible beating in the Fraunhofer envelope modulating the SQUID oscillations. The two different configurations of the junctions result in $W/L = 2.5$ for the narrow junction and $W/L = 8.5$ for the wide one. The inset of Fig. S7(b) shows that the SQUID modulation amplitude is suppressed for magnetic fields of $B = 3.0 \pm 0.5\,mT$, $B = 6.0 \pm 0.5\,mT$, and $B = 9.5 \pm 1\,mT$. For the first two field values, the total supercurrent is not zero, suggesting that for these field values only one junction carries a negligible supercurrent because it is at a Fraunhofer minimum. Instead, total suppression of the supercurrent of the whole SQUID occurs for $B = 9.5\,mT$. These observations suggest that the resonant features at the smaller field values can be attributed to the larger junction, while the feature at $9.5\,mT$ is attributed to the Fraunhofer minimum of both junctions. This is consistent with the fact that the ratio $A_{JJ,1}/A_{JJ,2}$ has almost an integer value and that the first minimum of the smaller junction corresponds approximately to the third minimum of the larger junction. Converting these magnetic fields to areas leads to $A_{eff} = 1.55 \cdot A_{geo}$ for both junctions, which we attribute to the flux-focusing effect.

### S9. MAGNETOMETER RESPONSE FOR THE SYMMETRIC SQUID

In the main text, the dissipative response of the asymmetric device was presented. For comparison, in this section we report details regarding the dissipative response of the symmetric SQUID. Fig. S8(a) shows the $V-B$ curves of the device at $V_{bg} = 20\,V$, for different values of the current bias. Tuning $I_{bias}$ to $\sim 50\,nA$ allows to reach a voltage drop modulation of $\Delta V \sim 20\,\mu V$, similar to the asymmetric device. This corresponds to an optimal voltage transfer function $V_\Phi = 0.5\,mV/\Phi_0$ at $V_{bg} = 20\,V$ (panel (b)), which is gate-tunable and presents its maximum value for $V_{bg}$ above $15\,V$ (panel (c)).

Using the value of the preamplifier input voltage noise, $4\,nV/\sqrt{Hz}$, we find that the magnetic flux noise at low frequency is $S_\Phi^{1/2} = 8.0\,\mu\Phi_0/\sqrt{Hz}$ at $T = 350\,mK$.

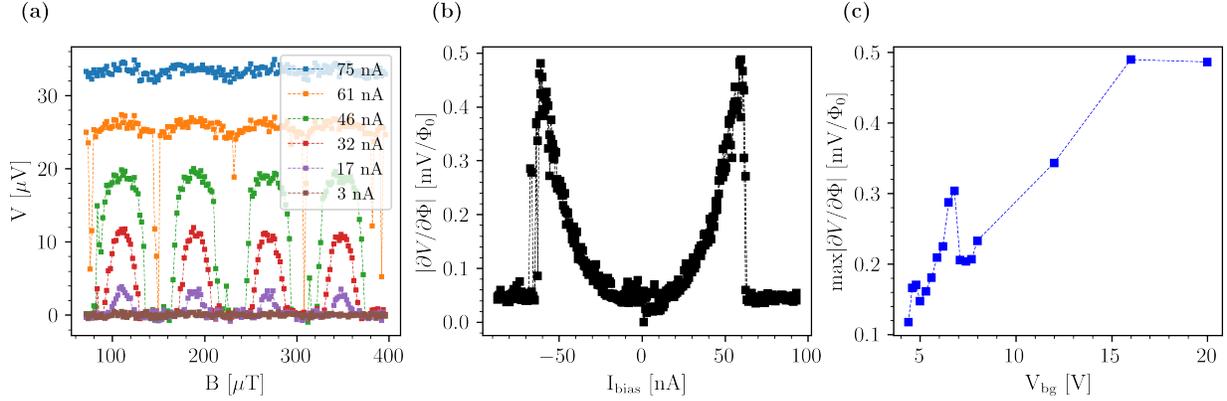

FIG. S8. (a) $V - B$ characteristics for different values of current bias. $T = 350\,\text{mK}, V_{\text{bg}} = 20\,\text{V}$. (b) Corresponding voltage responsivity as a function of current bias. (c) Maximum responsivity as a function of back-gate voltage.